\documentclass[apj]{emulateapj}
\usepackage{amsmath,amssymb}
\usepackage{mathptmx}
\usepackage{graphicx}
\usepackage{datetime}

\usepackage[backref,breaklinks,colorlinks,citecolor=blue,linkcolor=blue]{hyperref}

\usepackage[all]{hypcap} 
\def\sectionautorefname~#1\null{\S#1\null}
\def\subsectionautorefname~#1\null{\S\S#1\null}
\def\subsubsectionautorefname~#1\null{\S\S\S#1\null}

\newcommand{\ltsim}{\raisebox{-.5ex}{$\;\stackrel{<}{\sim}\;$}}
\newcommand{\gtsim}{\raisebox{-.5ex}{$\;\stackrel{>}{\sim}\;$}}

\newcommand{\todo}{\ifmmode {\Huge \bullet} \else {\Huge$\bullet$}\fi}
\newcommand{\til}{\ifmmode \sim \else $\sim$\fi}
\newcommand{\tido}{\ifmmode {{\bf\bullet}} \else {\bf$\bullet$}\fi}

\newcommand{\mic}	{\ifmmode {\rm \mu m} \else $\mu$m\fi}
\newcommand{\kms}	{\ifmmode {\rm km\,s}^{-1} \else km\,s$^{-1}$\fi}
\newcommand{\ergs}	{\ifmmode {\rm erg\,s}^{-1} \else erg s$^{-1}$\fi}
\newcommand{\Msun}{\ifmmode M_{\odot} \else $M_{\odot}$\fi}
\newcommand{\mpyr}{\ifmmode \Msun\,{\rm yr}^{-1} \else $\Msun\,{\rm yr}^{-1}$\fi}
\newcommand{\Msol}{\Msun}

\newcommand{  \Halpha   }{\ifmmode {\rm H}\alpha \else H$\alpha$\fi}
\newcommand{  \ha   	}{\ifmmode {\rm H}\alpha \else H$\alpha$\fi}
\newcommand{  \Hbeta    }{\ifmmode {\rm H}\beta \else H$\beta$\fi}
\newcommand{  \hb    	}{\ifmmode {\rm H}\beta \else H$\beta$\fi}
\newcommand{  \CIV      }{\ifmmode {\rm C}\,\textsc{iv}\,\lambda1549 \else C\,\textsc{iv}\,$\lambda1549$\fi}
\newcommand{  \mgii     }{\ifmmode {\rm Mg}\,\textsc{ii} \else Mg\,\textsc{ii}\fi}
\newcommand{  \MgII     }{\ifmmode {\rm Mg}\,\textsc{ii}\,\lambda2798 \else Mg\,\textsc{ii}\,$\lambda2798$\fi}

\newcommand{ \fwhb  }{\ifmmode {\rm FWHM}\left(\hb\right) \else FWHM(\hb)\fi}
\newcommand{ \fwmg  }{\ifmmode {\rm FWHM}\left(\mgii\right) \else FWHM(\mgii)\fi}
\newcommand{  \lamLlam  }{\ifmmode \lambda L_{\lambda} \else $\lambda L_{\lambda}$\fi}
\newcommand{  \Lop      }{\ifmmode L_{5100} \else $L_{5100}$\fi}
\newcommand{\fbol}{\ifmmode f_{\rm bol} \else $f_{\rm bol}$\fi}
\newcommand{\fbolwv}{\ifmmode f_{\rm bol}\left(\lambda\right) \else $f_{\rm bol}\left(\lambda\right)$\fi}
\newcommand{\fbolopt}{\ifmmode f_{\rm bol}\left(5100{\rm \AA}\right) \else $f_{\rm bol}\left(5100{\rm \AA}\right)$\fi}
\newcommand{\fbolthree}{\ifmmode f_{\rm bol}\left(3000{\rm \AA}\right) \else $f_{\rm bol}\left(3000{\rm \AA}\right)$\fi}

\newcommand{  \mbh      }{\ifmmode M_{\rm BH} \else $M_{\rm BH}$\fi}
\newcommand{  \lledd    }{\ifmmode L/L_{\rm Edd} \else $L/L_{\rm Edd}$\fi}
\newcommand{  \mmedd    }{\ifmmode \dot{m}/\dot{m}_{\rm \,Edd} \else $\dot{m}/\dot{m}_{\rm \,Edd}$\fi}
\newcommand{  \Lbol     }{\ifmmode L_{\rm bol} \else $L_{\rm bol}$\fi}
\newcommand{  \Lthree   }{\ifmmode L_{3000} \else $L_{3000}$\fi}

\newcommand{  \tacc    }{\ifmmode t_{\rm acc} \else $t_{\rm acc}$\fi}
\newcommand{  \tgrow     }{\ifmmode t_{\rm growth} \else $t_{\rm growth}$\fi}
\newcommand{  \tUni      }{\ifmmode t_{\rm U} \else $t_{\rm U}$\fi}
\newcommand{  \Mdotbh	}{\ifmmode \dot{M}_{\rm BH} \else $\dot{M}_{\rm BH}$\fi}
\newcommand{  \Mdotdisk	}{\ifmmode \dot{M}_{\rm disk} \else $\dot{M}_{\rm disk}$\fi}
\newcommand{  \Mdotacc	}{\ifmmode \dot{M}_{\rm acc} \else $\dot{M}_{\rm acc}$\fi}
\newcommand{  \Mdotthin	}{\ifmmode \dot{M}_{\rm thin} \else $\dot{M}_{\rm thin}$\fi}

\newcommand{  \as	}{\ifmmode a_{\rm *} 		\else $a_{\rm *}$\fi}
\newcommand{  \avec	}{\ifmmode \vec{a}_{\rm *} 	\else $\vec{a}_{\rm *}$\fi}
\newcommand{  \re	}{\ifmmode \eta      	\else $\eta$\fi}
\newcommand{  \mseed    }{\ifmmode M_{\rm seed} \else $M_{\rm seed}$\fi}

\newcommand{  \spitzer }  {{\it Spitzer}}
\newcommand{  \Spitzer }  {{\it Spitzer}}
\newcommand{  \WISE    }  {{\it WISE}}

\newcommand{\kband}{\textit{K}-band}

\newcommand{  \Ntot   }{20}
\newcommand{  \Ngood  }{20}
\newcommand{  \Nmbhraw}{35}

\newcommand{  \NSpitzer}{9}
\newcommand{  \NWISE  }{11}

\newcommand{  \ulas      }{ULAS\,J1120}
\newcommand{  \joneone   }{SDSS\,J1148}
\newcommand{  \jzeroone  }{SDSS\,J0100} %
\newcommand{  \jzerothree}{SDSS\,J0306} 

\newcommand{\zsix}{\ifmmode z \simeq 6 \else $z \simeq 6$\fi}

\newcommand{\Lspone}{\ifmmode L_{\rm 3.6} \else $L_{\rm 3.6}$\fi}
\newcommand{\Lsptwo}{\ifmmode L_{\rm 4.5} \else $L_{\rm 4.5}$\fi}

\shorttitle{Radiative Efficiencies of $z\gtsim6$ Quasars}
\shortauthors{Trakhtenbrot, Volonteri, \& Natarajan}

\begin{document}

\title{On the Accretion Rates and Radiative Efficiencies of the Highest Redshift Quasars}

\author{
Benny Trakhtenbrot\altaffilmark{1,4}, 
Marta Volonteri\altaffilmark{2}, and
Priyamvada Natarajan\altaffilmark{3}
}

\altaffiltext{1}{Institute for Astronomy, Department of Physics, ETH Zurich, Wolfgang-Pauli-Strasse 27, CH-8093 Zurich, Switzerland}
\altaffiltext{2}{Institut d'Astrophysique de Paris, UPMC et CNRS, UMR 7095, 98 bis bd Arago, F-75014 Paris, France}
\altaffiltext{3}{Department of Astronomy, Yale University, 260 Whitney Avenue, New Haven, CT 06511, USA}
\altaffiltext{4}{Zwicky postdoctoral fellow}

\slugcomment{Published as \textbf{ ApJL, 836, L1} }

\email{benny.trakhtenbrot@phys.ethz.ch}

\begin{abstract}
We estimate the accretion rates onto the supermassive black holes that power \Ngood\ of the highest-redshift quasars, at $z\gtrsim5.8$, including the quasar with the highest redshift known to date -- \ulas\ at $z=7.09$.
The analysis is based on the observed (rest-frame) optical luminosities and reliable ``virial'' estimates of the BH masses of the quasars, and 
utilizes scaling relations derived from thin accretion disk theory. 
The mass accretion rates through the postulated disks cover a wide range, $\Mdotdisk\simeq4-190\,\mpyr$, with most of the objects (80\%) having $\Mdotdisk\simeq10-65\,\mpyr$, 
confirming the Eddington-limited nature of the accretion flows.
By combining our estimates of \Mdotdisk\ with conservative, lower limits on the bolometric luminosities of the quasars, we investigate which alternative values of \re\ best account for all the available data.
We find that the vast majority of quasars ($\sim85\%$) can be explained with radiative efficiencies in the range $\re\simeq0.03-0.3$, with a median value close to the commonly assumed $\re=0.1$.
Within this range, we obtain conservative estimates of $\eta \gtrsim 0.14$ for \ulas\ and \jzeroone\ (at $z=6.3$), and of $\gtrsim0.19$ for \joneone\ (at $z=6.41$; assuming their BH masses are accurate).
The implied accretion timescales are generally in the range $\tacc\equiv\mbh/\Mdotbh\simeq 0.1 - 1$ Gyr,  suggesting that most quasars could have had $\sim1-10$ mass $e$-foldings since BH seed formation.
Our analysis therefore demonstrates that the available luminosities and masses for the highest-redshift quasars can be explained self-consistently within the thin, radiatively efficient accretion disk paradigm.
Episodes of radiatively \emph{inefficient}, ``super-critical'' accretion may have occurred at significantly earlier epochs (i.e., $z\gtrsim10$).
\end{abstract}

\section{Introduction}
\label{sec:intro}

The existence of luminous quasars as early as $z\sim6-7$ suggests that supermassive black holes (SMBHs) with masses of order $\mbh\sim10^{9}\,\Msol$ were in place less than 1 Gyr after the Big Bang. 
This is explicitly shown by observations that trace the gas dynamics in the close vicinity of the accreting SMBHs \cite[][]{Kurk2007,Willott2010_MBH,DeRosa2011,Trakhtenbrot2011,DeRosa2014}.
Such masses require continuous, exponential growth at the Eddington limit, $\lledd=1$, for almost the whole age of the universe at that time, and these conditions may not be necessarily ubiquitous among early SMBHs \cite[e.g.,][]{Treister2013,Habouzit2016,Trakhtenbrot2016_COSMOSFIRE_MBH,VolonteriReines2016_MM}.

The ability of SMBHs to grow to $\mbh\sim10^{9}\,\Msol$ depends, critically, on the radiative efficiency of the accretion flow -- defined as $\eta\equiv\Lbol/\Mdotacc c^2$, where \Mdotacc\ is the mass inflow rate and \Lbol\ is the emerging bolometric luminosity.
Most calculations of early BH growth assume a universal value of $\re\simeq0.1$, relying on the ensemble properties of quasars and relic SMBHs across all cosmic epochs. 

In reality, however, the role of \re\ in early BH growth is more complex.
Individual systems should have various \re, as indeed suggested by the observed range of BH spins (see, e.g., the review by \citealt{Reynolds2014_spin_rev} and also \citealt{DavisLaor2011_AD,Trakhtenbrot2014_hiz_spin,Capellupo2016_XS_pap3}).
The value of $\re\simeq0.1$ is within the range expected in optically thick, geometrically thin accretion disks, where $\re\sim0.04-0.32$, depending on the BH spin.
However, we recall that $\eta$ would be much lower for geometrically thick accretion disks, such as for advection-dominated flows, characteristic of significantly sub-Eddington accretion \cite[e.g.,][]{Narayan1995_ADAF}, or ``super-critical'' accretion flows \cite[e.g.,][]{Paczynsky1980_thick_ADs}. Additionally, in these regimes, the luminosity is not proportional to the accretion rate, but to the accretion rate squared and the logarithm of the accretion rate, respectively. 

Importantly, the relevance of either low-\re\ mechanisms or the assumption of a universal $\re=0.1$ to the observed population of the earliest known quasars, are not yet established.

In this \emph{Letter} we use insights from thin accretion disk theory and basic observables of some of the highest-redshift quasars known to date, at \zsix, to investigate the mass accretion rates and the corresponding radiative efficiencies powering these systems.
This work assumes a cosmological model with $\Omega_{\Lambda}=0.7$, $\Omega_{\rm M}=0.3$, and $H_{0}=70\,\kms\,{\rm Mpc}^{-1}$.

\section{Method, Sample, and Data}
\label{sec:data_analysis}

The goal of the present study is to test whether the currently available data for the highest-redshift quasars known to date can be self-consistently explained within the thin accretion disk model and to estimate the corresponding accretion rates (\Mdotdisk) and radiative efficiencies (\re). 
The method we use is based on two fundamental assumptions: 
that the SMBHs we study are powered by thin accretion disks and that their masses are reliably known.
In thin-disk accretion flows, the rest-frame optical continuum emission ($\lambda_{\rm rest}\gtrsim4500$ \AA), originating primarily from the outer disk, follows a power-law form, $L_\nu \propto \left(\mbh\, \Mdotdisk\right)^{2/3}\,\nu^{1/3}$ \cite[see, e.g.,][and references therein]{DavisLaor2011_AD}. 
Rewriting \Mdotdisk\ in terms of \mbh\ and the (monochromatic) continuum luminosity along this power-law tail provides
\begin{equation}
 \Mdotdisk \simeq 2.4\, \left(\frac{\lambda L_{\lambda,45}}{\cos i}\right)^{3/2}\,\left(\frac{\lambda}{5100{\rm \AA}}\right)^2 \, M_{8}^{-1} \,\,\, \mpyr \,\, ,
 \label{eq:mdot_ad}
\end{equation} 
where $\lambda$ is the (rest-frame) wavelength at which the continuum is measured, 
$\lambda L_{\lambda,45}\equiv\lamLlam/10^{45}\,\ergs$ denotes the monochromatic luminosity, 
and $M_{8}\equiv\mbh/10^{8}\,\Msun$. 
$\cos i$ represents the inclination between the line of sight and the polar axis of the disk (here we adopt $\cos i=0.8$, as appropriate for broad-line quasars).
The derivation of this expression is discussed in detail in, e.g., \cite{DavisLaor2011_AD} and \cite{NetzerTrakht2014_slim}.
This approach was used in several recent studies of accretion flows for samples of quasars to $z\simeq3.5$ \cite[e.g.,][]{BianZhao2003_spin,DavisLaor2011_AD,Wu2013_eff,Trakhtenbrot2014_hiz_spin}.
At $z\gtsim6$ the estimation of \Mdotdisk\ through \autoref{eq:mdot_ad} necessitates flux measurements at $\sim3.5-10$ \mic, and \kband\ spectroscopy of the \MgII\ broad emission line \cite[for \mbh\ estimation; see, e.g., ][hereafter TN12, and references therein]{TrakhtNetzer2012_Mg2}.
Here, we focus only on those $z\gtsim6$ quasars for which such data are publicly available.
Compiling all the \zsix\ quasars for which \mgii-based estimates of \mbh\ are available from NIR spectroscopy, we find \Nmbhraw\ objects in the studies of 
\cite{Iwamuro2004_z6}, \cite{Jiang2007}, \cite{Kurk2007}, \cite{Kurk2009}, \cite{Willott2010_MBH}, \cite{Venemans2013_VIKING_z6}, \cite{DeRosa2014}, and 
\cite{Venemans2015_PS1_z6}.
We additionally include the well-studied $z\simeq6.4$ quasar SDSS J1148 \cite[from][]{Barth2003_J1148}, and the 
highest-redshift quasar known to date, \ulas\ \cite[$z=7.085$;][]{Mortlock2011_z7}
We finally include the extremely luminous and massive quasars J0100+2802 \cite[$z=6.3$;][]{Wu2015_z6_nature} and J0306+1853 \cite[$z=5.363$;][]{Wang2015_z5_hiM}.
Although J0306 is at a lower redshift, we include it to test if the MIR-based high-redshift quasar selection methods may be unveiling a distinct population of SMBHs. 
Throughout this work, we highlight the results obtained for these four quasars of interest, but note here that they should be viewed as part of an ensemble of quasars.

For some of the quasars there are multiple published NIR spectra and/or \mgii\ profile measurements. 
Whenever possible, we have consistently used the detailed measurements performed by \cite{DeRosa2011}.
These replace the measurements provided by \cite{Iwamuro2004_z6}, \cite{Jiang2007}, \citet[][except for J0836]{Kurk2007}, and \cite{Kurk2009}.
For sources with multiple sets of measurements in \cite{DeRosa2011}, we selected those with the smaller uncertainties on \fwmg.
Using the measurements of \Lthree\ and \fwmg\ reported in the selected studies (and adjusted for our chosen cosmological model), we re-calculated all \mbh\ estimates following the prescription of TN12 \cite[see also][]{Shen_dr7_cat_2011}. 
These accurate estimates of \mbh\ are known to carry systematic uncertainties of up to 0.5 dex (TN12 and references therein).

We then compiled rest-frame optical photometric data, obtained with the \spitzer\ and \WISE\ IR space telescopes.
For nine of the quasars, we use \spitzer/IRAC data reported in the studies of 
\cite{Jiang2006_Spitzer}, \cite{Leipski2014}, and \citet[for \ulas]{Barnett2014_ULAS1120_SED}.
Whenever possible, we used the \cite{Leipski2014} measurements for homogeneity.
For \NWISE\ quasars we use \WISE\ measurements in the W1 and W2 bands, obtained by cross-matching our sample with the {\it AllWISE} data release, within 5\arcsec\ of the optical coordinates of the sources.
In all cases where both \spitzer\ and \WISE\ measurements are available, we preferred the higher spatial resolution and sensitivity \spitzer\ data.
The first two bands of the IRAC or \WISE\ cameras have effective wavelengths of roughly 3.6 and 4.5 \mic. 
For our sample's redshift range, these correspond to rest-frame wavelengths of about 4850-5340 and 6065-6675 \AA, respectively.
We verified that none of our sources is affected by blending with neighboring \WISE\ sources.
All \spitzer\ and \WISE\ measurements were converted to flux densities using standard procedures \cite[i.e.,][]{Wright2010_WISE,Jarrett2011_WISE_Spitzer}.
Monochromatic luminosities were calculated assuming the \mgii-based redshifts reported in the aforementioned NIR studies (see \autoref{tab:lums_masses}).

Our final sample includes \Ntot\ quasars with reliable estimates of \mbh, \NSpitzer\ with \spitzer\ data and the remaining \NWISE\ with \WISE\ data.
We verified that none of the choices we made in compiling the data set has significant effects on our results.
The heterogeneous nature of our sample -- drawn from several surveys of varying depth, and our obvious focus on vigorously accreting SMBHs at this extremely high redshift regime, mean that our sample is most probably not representative of the entire population of (active) SMBHs at \zsix.

\begin{figure}[t!]
\includegraphics[angle=-00,width=0.48\textwidth]{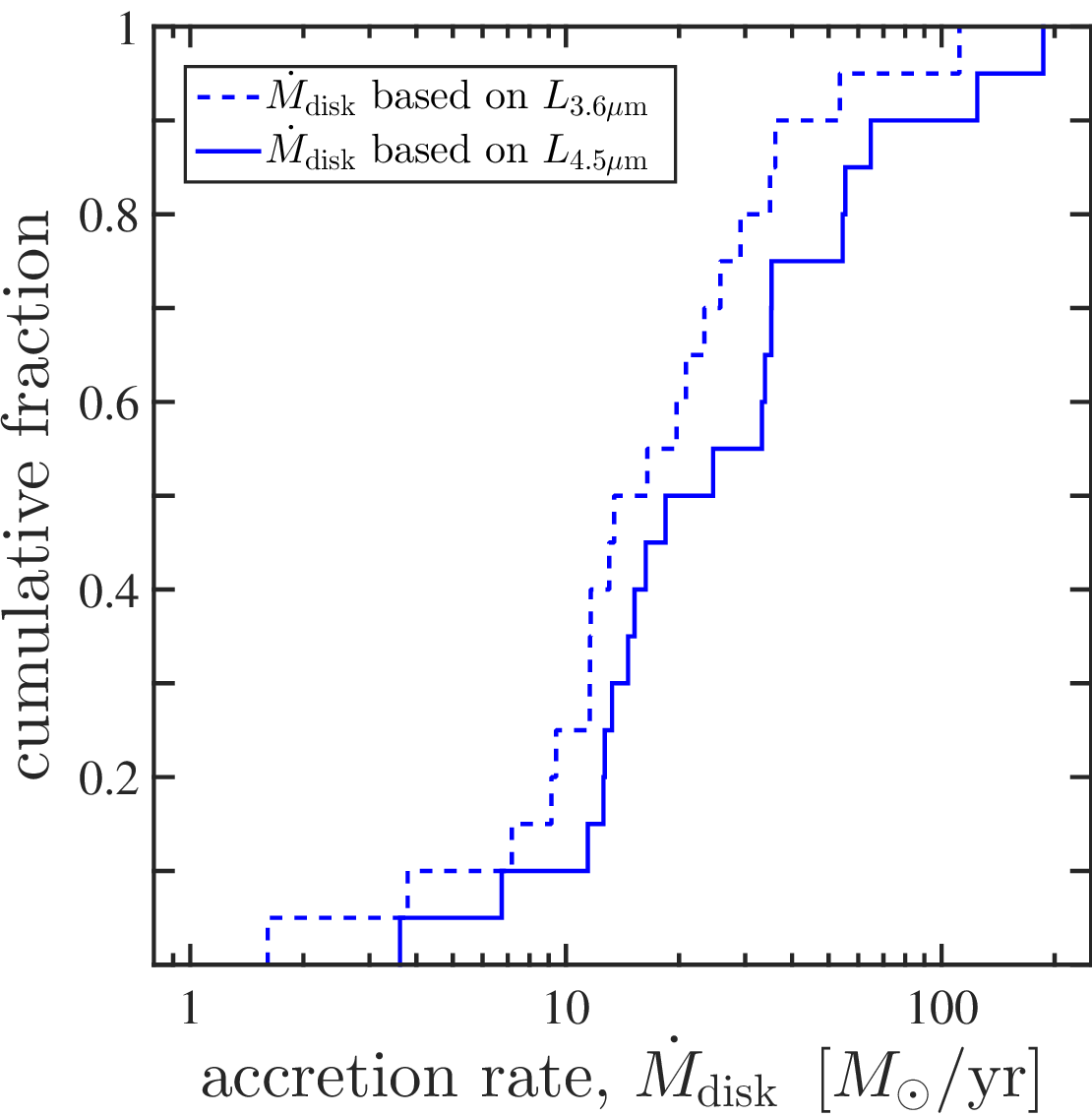} 
\caption{
Cumulative distribution function of our estimates of accretion rates through the disks, \Mdotdisk, based on \autoref{eq:mdot_ad}.
The dashed and solid lines represent the estimates based on \Lspone\ and \Lsptwo, respectively. 
The systematically higher \Lsptwo-based estimates of \Mdotdisk\ result in more conservative constraints on \re\ (i.e., lower values) and shorter accretion timescales.
}
\label{fig:Mdot_cdf}
\end{figure} 

%
We calculated \Mdotdisk\ for the \Ntot\ quasars through \autoref{eq:mdot_ad}, 
using the (re-calculated) \mgii-based \mbh\ estimates, and the monochromatic luminosities observed in either of the IR bands (i.e., $\sim$3.6 and 4.5 \mic, hereafter \Lspone\ and \Lsptwo, respectively).
We list all the quantities relevant for the present analysis, including the chosen sources of all measurements, in \autoref{tab:lums_masses}.

\section{Results}
\label{sec:results}

Figure~\ref{fig:Mdot_cdf} presents the derived accretion rates through the disks, \Mdotdisk, for all the sources in our sample, and based on the two different (rest-frame) optical luminosities.
The accretion rates we obtain using \Lsptwo\ are in the range $\Mdotdisk\simeq 3.6 - 187\,\mpyr$, with 16 of the quasars (80\%) having $\Mdotdisk\simeq10-65\,\mpyr$.
We note that some of the variance in \Mdotbh\ in our sample may be attributed to the significant systematic uncertainties in \mbh, and the form of \autoref{eq:mdot_ad}.
Notwithstanding this limitation, we obtain $\Mdotdisk=11.4\,\mpyr$ for \ulas, while for \joneone\ and \jzeroone\ we find $\Mdotdisk\simeq16.3$ and $54.6$ \mpyr, respectively. 
The accretion rate we find for \jzerothree\ is in excellent agreement with that of \jzeroone, which is expected given the very similar masses and continuum luminosities of the two quasars (both within 0.1 dex).
The \Lspone-based \Mdotdisk\ estimates are systematically lower than those based on \Lsptwo, by about 0.19 dex (median value).
This is probably due to the fact that the 4.5 \mic\ band includes the strong broad \Halpha\ line emission, which is expected to be stronger by a factor of $\sim$3 compared with the \Hbeta\ line, covered in the 3.6 \mic\ band data of most sources.
Moreover, the 3.6 \mic\ band is probing the continuum emission in a spectral regime where the power-law approximation may no longer be valid \cite[particularly at high \mbh; see, e.g.,][]{DavisLaor2011_AD,NetzerTrakht2014_slim}.
In what follows, we focus on the \Lsptwo-based estimates of \Mdotdisk, as these would result in more conservative constraints on \re\ (i.e., lower limits; see below).
%

\begin{figure}[t]
\includegraphics[angle=-00,width=0.49\textwidth]{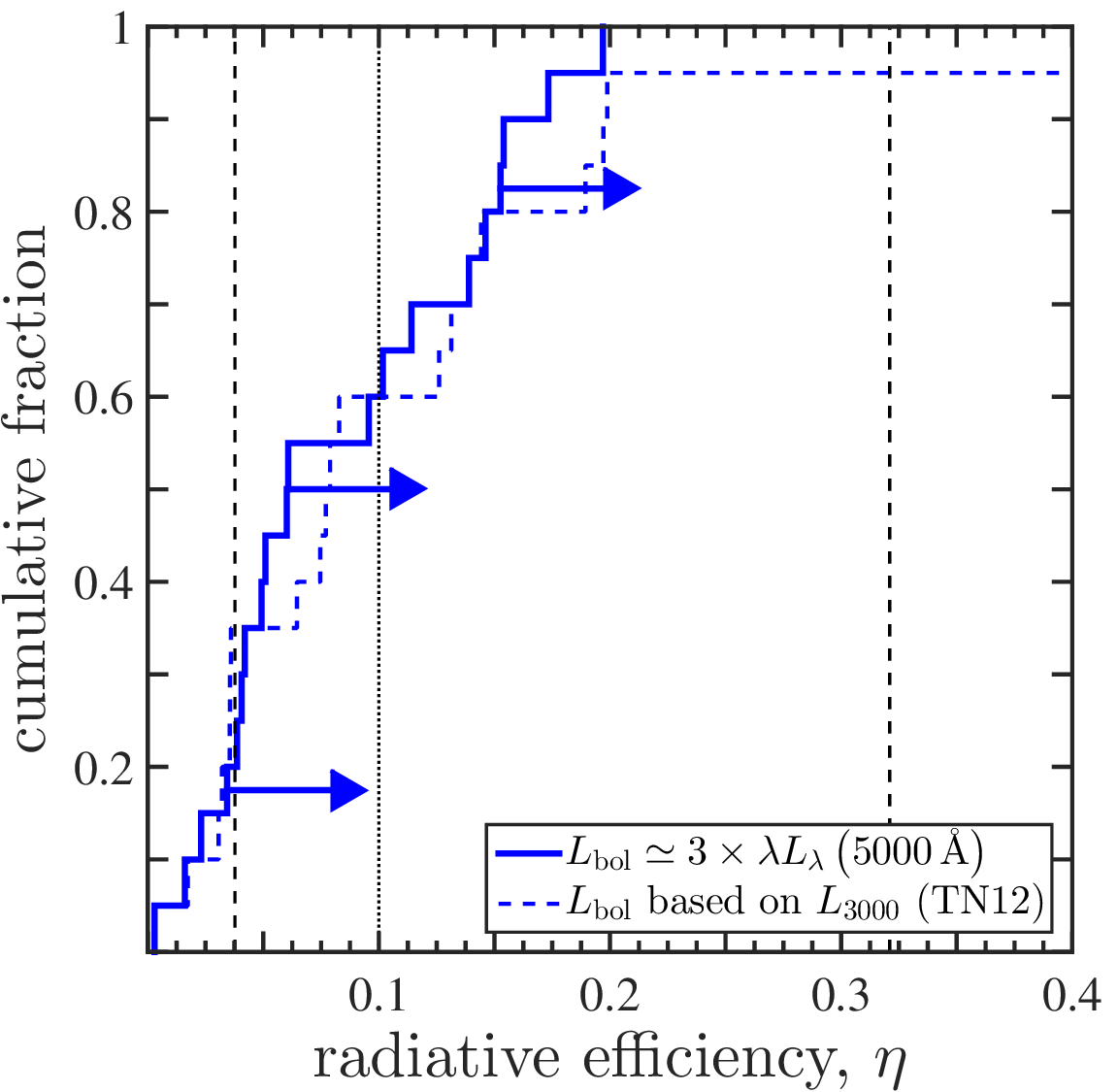} 
\caption{
Cumulative distribution function of radiative efficiency estimates, \re.
The solid line traces the values obtained using the conservative assumption of $\Lbol=3\times\Lspone$ (roughly $\Lbol\simeq3\times\Lop$) and the generally higher \Lsptwo-based estimates of \Mdotdisk.
The dashed line traces the \re\ estimates based on $\Lbol\left(\Lthree\right)$, using the TN12 bolometric corrections.
The vertical dashed lines mark range of radiative efficiencies expected for thin accretion disks around spinning BHs ($0.038\lesssim\re\lesssim 0.32$).
The dotted vertical line marks $\re=0.1$.
}
\label{fig:eta_cdf}
\end{figure} 

\begin{figure}[t]
\includegraphics[angle=-00,width=0.48\textwidth]{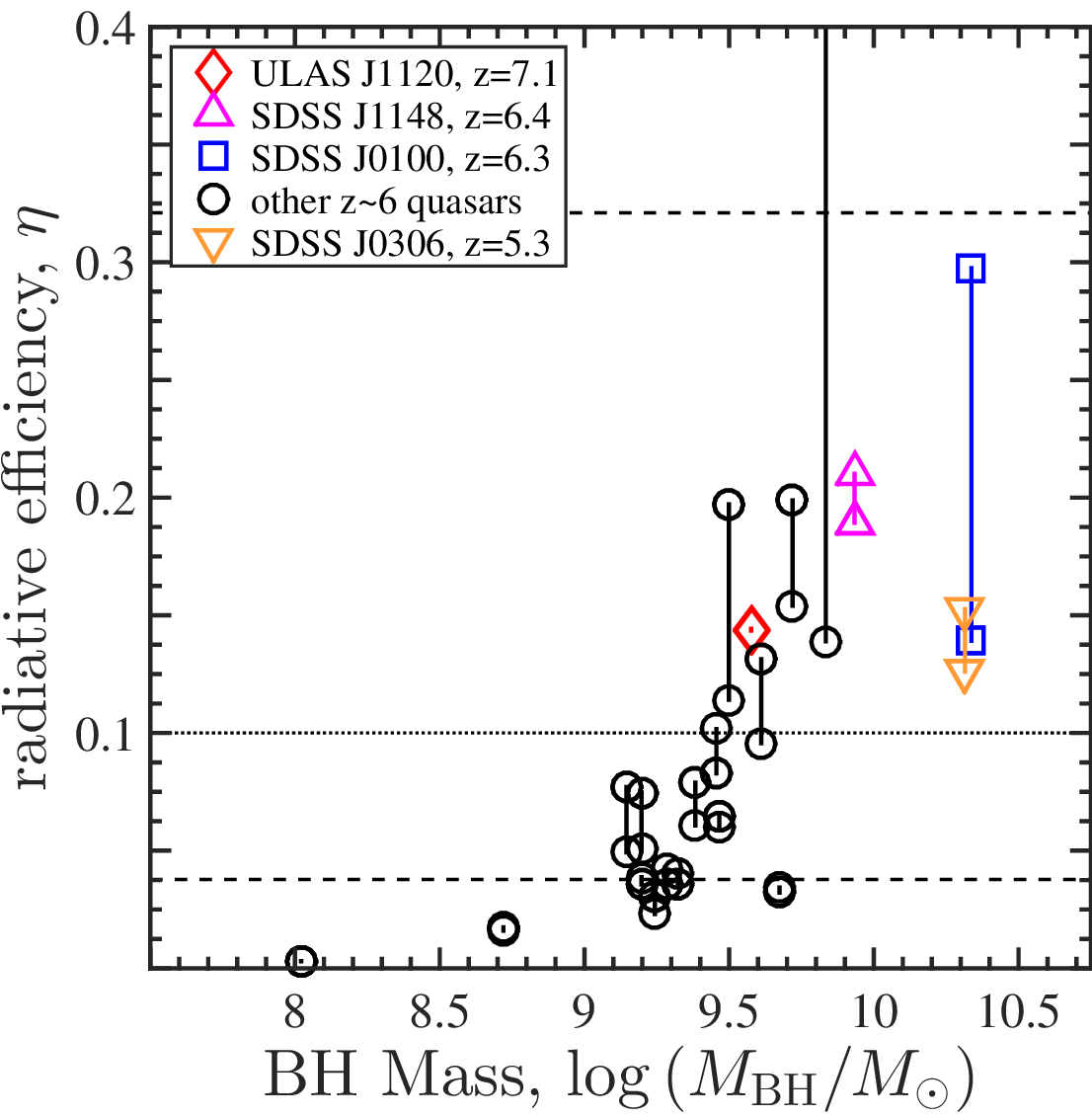} 
\caption{
Estimates of radiative efficiencies \re, vs. BH mass, \mbh, for the \Ntot\ quasars compiled in our sample.
For each quasar, we show two different estimates of \re, with the lower value being our most conservative estimate (based on $\Lbol=3\times\Lspone$\ and the \Lsptwo-based  \Mdotdisk\ estimates), and the higher value derived through the \Lthree-based \Lbol\ estimates (the TN12 bolometric corrections).
The real \re\ may be higher than what these two sets of estimates suggest.
The most massive BHs in our sample show high radiative efficiencies, $\re\gtrsim0.2$.
The apparent trend of increasing \re\ with increasing \mbh\ is likely driven by the form of \autoref{eq:mdot_ad}. We cannot rule out that SMBHs with $\mbh\gtrsim10^{10}\,\Msol$ and $\re \lesssim 0.1$ exist, but are not (yet) observed.
}
\label{fig:eta_vs_mbh}
\end{figure} 

\begin{figure*}[t!]
\includegraphics[width=0.98\textwidth,angle=-00]{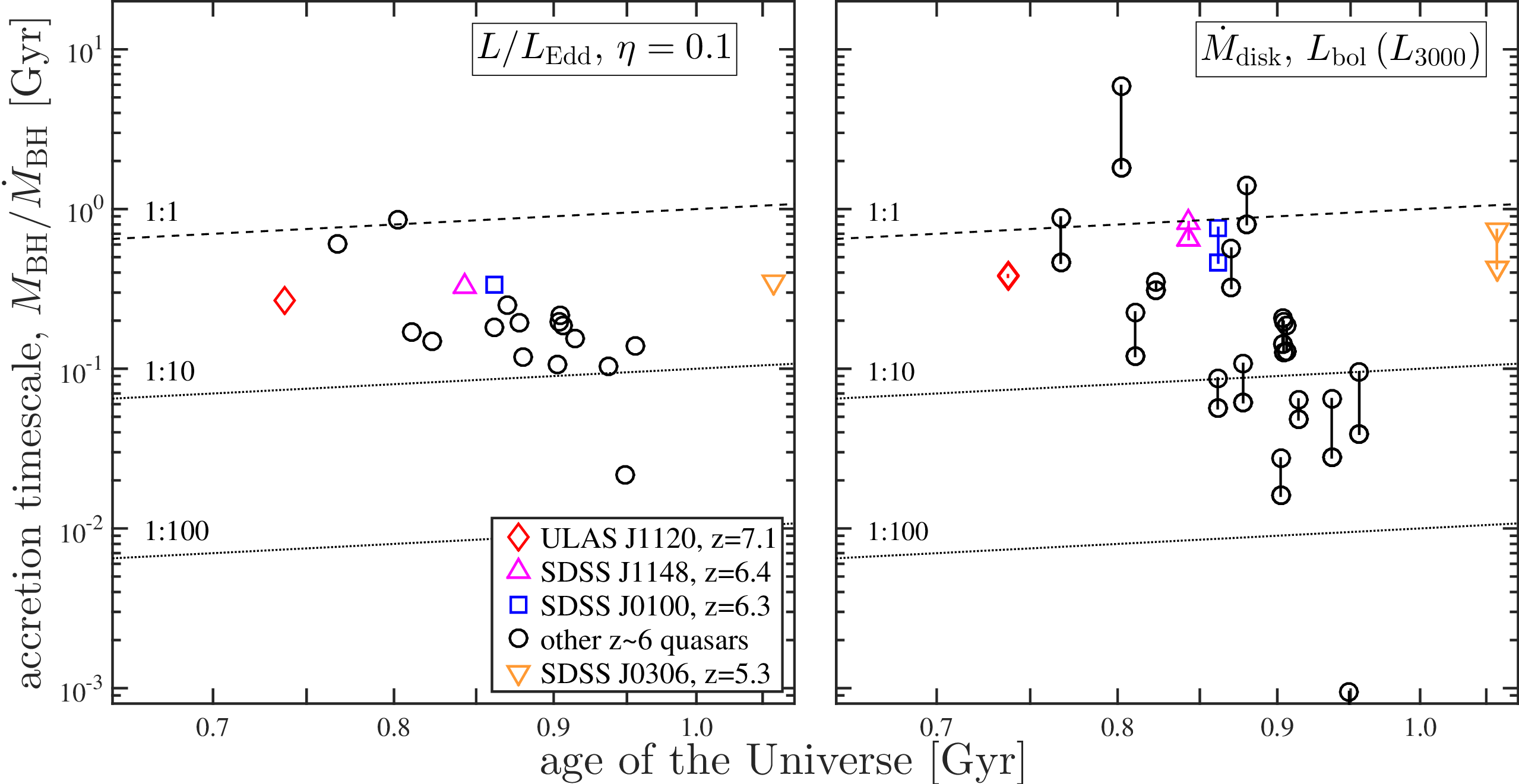} 
\caption{
Different estimates of BH accretion timescales, $\tacc=\mbh/\Mdotbh$, vs. cosmic epoch.
{\it Left:} ``standard'' \tacc\ timescale estimates, obtained from \lledd\ and assuming $\re=0.1$.
{\it Right:} \tacc\ estimates obtained from \Mdotdisk\ and the \Lthree-based estimates of \Lbol\ (i.e., equivalent to using our \re\ estimates; see \autoref{tab:lums_masses}). 
For each source, we plot the timescales obtained from both the \Lspone- and \Lsptwo-based estimates of \Mdotdisk.
}
\label{fig:tacc_vs_tuni_both}
\end{figure*} 

We next use the estimates of \Mdotdisk\ and the observed luminosities of our quasars to investigate the range of -- or lower limits on -- \re, which would be consistent with all the data available for the \zsix\ quasars, following $\eta=\Lbol/\Mdotdisk c^2$. 
Given the data in hand, \Lbol\ can only be estimated by utilizing bolometric corrections and monochromatic luminosities \cite[unlike the analysis of][]{DavisLaor2011_AD}.

We focus on conservative, lower limits on \re, which can be obtained assuming $\Lbol=3\times\Lspone$ ($\simeq3\times\lamLlam\left[4930 {\rm \AA}\right]$).
This choice is very similar to the one made in \cite{Trakhtenbrot2014_hiz_spin} ($\Lbol=3\times\lamLlam\left[5100{\rm \AA}\right]$).
We consider it to provide a lower limit on the real \Lbol\ since it reflects a bolometric correction that is much smaller, by at least a factor of $\sim$2,  than those used in many other studies of \mbh\ and \lledd\ in high-redshift quasars (see, e.g., \citealt{Runnoe2012}).
Since we seek to derive lower limits on \re, we further use the higher, \Lsptwo-based estimates of \Mdotdisk.
As $\re\propto 1/\Mdotdisk\propto\mbh$, our \re\ estimates inherit the systematic uncertainties on \mbh\ (see above).
\autoref{fig:eta_cdf} shows the cumulative distribution of the conservative constraints on \re\ we obtain, which are in the range $\re\gtrsim0.003-0.2$. 
Most quasars have lower limits on \re\ that are consistent with what is expected for thin disks, and only three quasars have lower limits on \re\ that are below 0.03.
For the three quasars of particular interest we find conservative lower limits of $\re\gtrsim0.14$, 0.17, and 0.2 (for  \ulas, \joneone, and \jzeroone, respectively). 
For \jzerothree\ we find $\re\gtrsim0.15$,  consistent with the \zsix\ quasars and particularly with \jzeroone, which has a very similar BH mass.

To further test the range of \re\  consistent with the data, we also calculated \Lbol\ using the \Lthree-dependent bolometric corrections of TN12, which for the sample in hand are in the range $\fbolthree\sim2-3.2$.
Coupling these \Lbol\ estimates with the \Lsptwo-based \Mdotdisk\ estimates, we obtain  \re\ estimates in the range of $\eta\simeq 0.003-0.44$ (see \autoref{tab:lums_masses} and the dashed line in \autoref{fig:eta_cdf}).
In this case, 12 of the \Ngood\ quasars (60\%) have \re\ within the  range of values expected for thin disks. 
Only two objects have $\eta<0.03$, and only one has $\eta>0.3$.
Focusing again on the three quasars mentioned above, we obtain $\re\simeq0.14$, 0.19, and 0.14, for \ulas, \joneone, and \jzeroone, respectively.
For \jzerothree\ at $z=5.36$ we find $\re\simeq0.13$, again highly consistent with \jzeroone.

We note that these latter \Lthree-based estimates of \Lbol\ and \re\ may also be considered conservative, as the TN12 bolometric corrections we use are, again, significantly lower than those commonly used for samples of high-$z$ quasars (by factors of $\sim2$; see, e.g., the compilation of \citealt{Runnoe2012}).
If we had instead used the \fbolthree\ suggested by \cite{Runnoe2012}, the resulting \re\ estimates would have been higher by $\sim30\%$.

\autoref{fig:eta_vs_mbh} presents our two sets of (conservative) \re\ estimates of \re\ against \mbh. 
The apparent trend of increasing \re\ with increasing \mbh\ is mainly driven by the explicit dependence of \autoref{eq:mdot_ad} on \mbh, and then on the fact that $\re\propto1/\Mdotdisk$ (at fixed luminosity).  
It is therefore expected that rest-frame UV-optical surveys of a given depth would miss high-\mbh\ but low-\re\ objects \cite[see also][]{Bertemes2016_SDSS_spin}.
In this context, extremely massive objects like \jzeroone\ and \jzerothree, which have $\mbh\gtrsim10^{10}\,\Msol$, may represent the high-\re\ end of a much larger population of high-mass BHs at $z\sim5-6$. 
A large enough number of such extreme, yet-to-be-observed objects may further constrain models of BH seed formation and early growth \cite[e.g.,][]{Agarwal2013_obese}.

We next turn to estimate the accretion timescales of the SMBHs powering the quasars, $\tacc\equiv\mbh/\Mdotbh$, which may be considered as the typical BH mass $e$-folding timescales.
Here, we derive and compare two sets of \tacc\ estimates made available by the data, and following two approaches for estimating $\Mdotbh=\left(1-\eta\right)\Mdotdisk$.
First, we follow the common procedure of using \lledd\ and a fixed radiative efficiency (i.e., $\re=0.1$), to derive accretion timescales through 
$\tacc\simeq0.4 \left(\re/1-\re\right)\,{\rm Gyr}$.
The \tacc\ estimates thus obtained are shown in the left panel of \autoref{fig:tacc_vs_tuni_both}, plotted against the age of the universe (at the observed epoch). 
Most objects have $\tacc\sim0.1-1$ Gyr, and could have had between $\sim1-10$ mass $e$-foldings of \mbh.
Second, we use our \Mdotdisk\ estimates, which provide $\Mdotbh=\left(1-\eta\right)\Mdotdisk$, and our conservative estimates of \re. 
Due to the dependence of \re\ on \Mdotdisk\ in our analysis, we note that these estimates can be expressed as $\tacc = \mbh / \left(\Mdotdisk - \Lbol/c^2\right)$. 
As before, we adopt the \Lthree-based estimates of \Lbol, and either the \Lspone- or \Lsptwo-based estimates of \Mdotdisk.
The accretion timescales we obtain through this procedure are presented in the right panel of \autoref{fig:tacc_vs_tuni_both}. 
The shorter, \Lspone-based timescales, are generally in the range of $\tacc\sim 0.01 - 2$ Gyr, with 18 of the quasars (90\%) having $\tacc\simeq0.03-0.8$ Gyr.
The extremely high-\Mdotdisk\ quasar J0005 ($\Mdotdisk>100\,\mpyr$) has the shortest accretion timescale, $\tacc \simeq 0.5$ Myr (cf.\ $\sim$0.02 Gyr obtained from \lledd).
For the three $z>6$ quasars of interest,  \ulas, \joneone, and \jzeroone, we find $\tacc=0.36$, 0.71 and 0.67 Gyr, respectively.
The ultramassive $z\simeq5.3$ quasar \jzerothree\ has $\tacc=0.66$ Gyr.
The \Mdotdisk-based estimates of \tacc\ suggest that some quasars could have had as many as 50 BH mass $e$-foldings.

We stress, again, that the trends of decreasing \tacc\ with epoch seen in \autoref{fig:tacc_vs_tuni_both} are due to the way the original samples of quasars were selected and identified (i.e., their brightness and luminosity), and the form of the prescriptions we use here.

The two sets of \tacc\ estimates are generally in good agreement, with differences being within a factor of $\sim$2 for 14 (70\%) of our quasars, and even within a factor of $\sim$1.5 for 8 (40\%) of the quasars. 
The lack of a significant systematic offset between the two timescale estimates reflects the fact that our \re\ estimates bracket the ``standard'' value of $\re=0.1$.
We conclude that for our sample of \zsix\ quasars, the simpler \lledd-based growth timescale estimates are broadly consistent with those derived through our more elaborate approach.
%

\section{Discussion and Conclusion}
\label{sec:discussion}

The main point of our analysis is to investigate whether the data available for some of the highest-redshift quasars known to date can be accounted for, self-consistently, within the generic model of a radiatively efficient, thin accretion disk. 

We found that the accretion rates through the postulated thin disks are generally in the range of $\Mdotdisk\sim1-100\,\mpyr$. 
These accretion rates are consistent with the systems being Eddington-limited, at the observed epoch.
However, if one assumes that these accretion rates were sustained at earlier epochs, when the BH masses were considerably lower, this would imply super-Eddington accretion rates, which may be sustained under certain gas configurations, and lead to a fast buildup of BH mass \cite[e.g.,][]{Volonteri2015_superEdd}.

We showed that the available data for most of the \zsix\ quasars can be explained with conservative estimates (lower limits) on the radiative efficiencies that are in the range $0.04 \lesssim \re \lesssim 0.32$ -- that is, within the range expected for accretion through a thin disk onto rotating BHs. Our more conservative estimates suggest $\re \gtsim 0.05$ for most objects. 
Thus, it appears that all the data available for quasars at $z\gtsim5.8$ are consistent with such radiatively efficient accretion flows.
Moreover, since our analysis provides lower limits on \re, it is possible that the real \re\ of the quasars under study would differ substantially with the expectations of radiatively \emph{inefficient} accretion flow models.

We stress that this result is \emph{not} a trivial consequence of the  observables and methodology we adopt here (i.e., \autoref{eq:mdot_ad}) . 
For example, the study of \cite{Trakhtenbrot2014_hiz_spin} -- applying the same methods as the ones used here to a sample of high-\mbh\ quasars at $1.5\lesssim z\lesssim3.5$ -- found extremely high values of \re, which in many cases exceeded $\re\simeq0.4$.
Among the conservative \re\ estimates, we find $\re\gtsim 0.15$ for \ulas\ -- the highest-redshift quasar known to date ($z=7.1$), and $\re\gtsim0.2$ for \joneone\ (at $z=6.4$). 
On the other hand, many other \zsix\ quasars have $0.05\ltsim\re\ltsim0.1$ -- below the standard, universal radiative efficiency assumed in many studies of the AGN population.
The limited size of our sample prevents us from determining whether these \re\ estimates represent the scatter within the quasar population, or only trace a few extreme cases.

The accretion timescales of the SMBHs under study, derived assuming the \Mdotdisk\ and \re\ estimates, are consistent with those derived from \lledd\ and the universal $\re=0.1$ assumption, allowing for $\sim1-100$ mass $e$-foldings.
This further justifies the usage of the simpler \tacc\ estimates in cases where only \lledd\ is available (i.e., when rest-frame optical luminosities are unavailable). 
However, this assumption would naturally neglect the fact that any population of accreting SMBHs is expected to have a range of \re.

Within the standard thin-disk framework, radiative efficiencies are closely linked to BH spin, \as\ (in normalized units). 
The range of \re\ we find corresponds to the entire possible range of $-1 \leq \as \leq 1$, and the typical (median) \re\ of our sample corresponds to $\as\simeq0.7$ -- again consistent with what is expected from the assumption of a universal $\re=0.1$.
As the quasars of interest in our sample (\ulas, \joneone, and \jzeroone) have $\re>0.1$, they correspond to rather high spins, $\as\gtsim 0.9$.
These, in turn, are consistent with what is found for low-redshift, low-luminosity AGNs \cite[][and references therein]{Reynolds2014_spin_rev} and for higher-luminosity, higher-\mbh\ quasars at higher redshifts \cite[][]{Reis2014_spin,Trakhtenbrot2014_hiz_spin,Capellupo2016_XS_pap3}.
This supports a scenario in which the \zsix\ SMBHs grew through coherent accretion flows \cite[e.g.,][and references therein]{Dotti2013,Volonteri2013_spin}. 
This ``spin up'' scenario appears highly plausible, given the high duty cycle of accretion required for the fast BH growth at $z>6$.

Several recent studies highlighted the possibility that \zsix\ quasars could have grown through ``super-critical'' accretion, to reach their high BH masses \cite[][]{VolonteriRees2005,Alexander2014_seeds,Volonteri2015_superEdd}.
The more extreme cases of such accretion, in slim disks, may result in $\mmedd\sim100$ and $\re \sim0.01$, essentially regardless of the BH spin 
(see, e.g., \citealt{Sadowski2009_slim,Madau2014_supEdd,McKinney2014_supEdd_sim,Volonteri2015_superEdd}, but see \citealt{McKinney2015_supEdd_eff}).
Our analysis confirms that such super-critical growth episodes should have occurred, if at all, in the yet earlier universe, when BHs were smaller, to alleviate the requirement of continuous growth, which may not be realistic.

The data currently available for the highest redshift quasars, namely, in the (rest-frame) UV and optical, as well as the data that may become available in the foreseeable future, cannot \emph{directly} distinguish between radiatively efficient and inefficient accretion. 
Models of such flows require further study, with an emphasis on the observables that are relevant for faint, $z>6$ sources. 
For instance, super-critical episodes are likely accompanied by the production of powerful jets  \cite[][]{McKinney2014_supEdd_sim,Sadowski2014_supEdd_sims,Sadowski2016_supEdd_sims}. 
At high-$z$ such jets should be detectable in X-rays rather than as extended radio sources \citep{Ghisellini2015}. 
This, together with the limitations present in the (rest-frame) UV regime due to IGM absorption, highlight the importance of the X-ray regime, where surveys of ever-increasing depth (e.g., in the CDF-S field) may provide key insights into the assembly of the earliest SMBHs.

\acknowledgements

We appreciate the feedback from the anonymous referee, which helped us to improve the manuscript.
We thank K.\ Schawinski  for beneficial discussions.
This work was performed in part at the Aspen Center for Physics, which is supported by National Science Foundation grant PHY-1066293. 
M.V.\ acknowledges funding from the European Research Council under the European Community's Seventh Framework Programme (FP7/2007-2013 Grant Agreement No.\ 614199, project ``BLACK'’).
P.N.\ acknowledges support from TCAN grant with award number 1332858 from the National Science Foundation.
This work made use of the MATLAB package for astronomy and astrophysics \cite[][]{Ofek2014_matlab}.

\newpage



\section*{}
\clearpage

\capstartfalse
\begin{deluxetable*}{lcccccccccccc}

\tablecolumns{12}
\tablewidth{0pt}
\tablecaption{Observed and Derived Properties \label{tab:lums_masses}}
\tablehead{
  \colhead{Object}  &
  \colhead{$z$ \tablenotemark{a}}  &
  \colhead{$\log\Lthree$ \tablenotemark{b}} &
  \colhead{\fwmg} &
  \colhead{NIR} &
  \colhead{$\log\mbh$ \tablenotemark{d}} &
  \colhead{$\log\Lspone$ \tablenotemark{e}} &
  \colhead{$\log\Lsptwo$ \tablenotemark{e}} &
  \colhead{MIR} &
  \colhead{$\Mdotdisk$ \tablenotemark{g}}  &
  \colhead{$\re$ \tablenotemark{h}}  \\
    &  & (\ergs) & (\kms) & Ref.\tablenotemark{c} & (\Msun) & (\ergs) & (\ergs) & Ref.\tablenotemark{f}  & (\mpyr) & }
\startdata
J1120$+$0641 & $7.097$ & $46.48$ & $4411$ & $2$ & $ 9.58$ & $46.50$ & $46.36$ & $3$ & $ 11.4$ & $0.144$ \\
J1148$+$5251 & $6.407$ & $46.79$ & $5352$ & $1$ & $ 9.93$ & $46.73$ & $46.65$ & $1$ & $ 16.3$ & $0.189$ \\
J0100$+$2802 & $6.3  $ & $47.50$ & $5130$ & $4$ & $10.34$ & $47.31$ & $47.24$ & $2$ & $ 54.6$ & $0.139$ \\
\hline \\ [-1.25ex]
J0306$+$1853 & $5.363$ & $47.31$ & $5722$ & $5$ & $10.32$ & $47.20$ & $47.15$ & $2$ & $ 55.4$ & $0.126$ \\
\hline \\ [-1.25ex]
J0050$+$3445 & $6.253$ & $46.55$ & $4360$ & $6$ & $ 9.61$ & $46.42$ & $46.37$ & $2$ & $ 14.6$ & $0.131$ \\
J0836$+$0054 & $5.81 $ & $46.93$ & $3600$ & $7$ & $ 9.67$ & $46.91$ & $47.01$ & $1$ & $124.4$ & $0.032$ \\
J0353$+$0104 & $6.072$ & $46.42$ & $3682$ & $1$ & $ 9.38$ & $46.32$ & $46.29$ & $1$ & $ 18.4$ & $0.079$ \\
J0842$+$1218 & $6.069$ & $46.47$ & $3931$ & $1$ & $ 9.47$ & $46.45$ & $46.43$ & $1$ & $ 24.6$ & $0.065$ \\
J2348$-$3054 & $6.889$ & $45.99$ & $5446$ & $2$ & $ 9.46$ & $46.11$ & $46.10$ & $2$ & $  6.7$ & $0.083$ \\
J0305$-$3150 & $6.605$ & $46.24$ & $3189$ & $2$ & $ 9.15$ & $46.07$ & $46.05$ & $2$ & $ 12.6$ & $0.077$ \\
P036$+$03    & $6.527$ & $46.68$ & $3500$ & $3$ & $ 9.50$ & $46.44$ & $46.28$ & $2$ & $ 12.7$ & $0.197$ \\
P338$+$29    & $6.658$ & $46.11$ & $6800$ & $3$ & $ 9.72$ & $46.02$ & $46.07$ & $2$ & $  3.6$ & $0.199$ \\
J0005$-$0006 & $5.844$ & $46.00$ & $1036$ & $1$ & $ 8.02$ & $46.02$ & $46.03$ & $1$ & $186.6$ & $0.003$ \\
J1411$+$1217 & $5.903$ & $46.57$ & $2824$ & $1$ & $ 9.24$ & $46.45$ & $46.55$ & $1$ & $ 64.8$ & $0.031$ \\
J1306$+$0356 & $6.017$ & $46.34$ & $3158$ & $1$ & $ 9.20$ & $46.39$ & $46.34$ & $1$ & $ 33.9$ & $0.036$ \\
J1630$+$4012 & $6.058$ & $46.25$ & $3366$ & $1$ & $ 9.20$ & $46.11$ & $46.07$ & $1$ & $ 13.3$ & $0.075$ \\
J0303$-$0019 & $6.079$ & $46.01$ & $2307$ & $1$ & $ 8.72$ & $46.00$ & $46.02$ & $1$ & $ 33.3$ & $0.017$ \\
J1623$+$3112 & $6.211$ & $46.36$ & $3587$ & $1$ & $ 9.32$ & $46.43$ & $46.45$ & $1$ & $ 35.2$ & $0.036$ \\
J1048$+$4637 & $6.198$ & $47.28$ & $3366$ & $1$ & $ 9.83$ & $46.60$ & $46.54$ & $1$ & $ 15.2$ & $0.444$ \\
J1030$+$0524 & $6.302$ & $46.36$ & $3449$ & $1$ & $ 9.28$ & $46.45$ & $46.43$ & $1$ & $ 35.2$ & $0.036$ 
\enddata
\tablenotetext{a}{Redshift measured from the best-fit model of the \mgii\ line.}
\tablenotetext{b}{Monochromatic luminosity (\lamLlam) at rest-frame wavelength of 3000 \AA, obtained from the best-fit model of the \mgii\ emission line complex.}
\tablenotetext{c}{References for NIR spectral analysis and \mgii\ measurements: (1) \cite{DeRosa2011}; (2) \cite{DeRosa2014}; (3) \cite{Venemans2015_PS1_z6}; (4) \cite{Wu2015_z6_nature}; 
(5) \cite{Wang2015_z5_hiM}; (6) \cite{Willott2010_MBH}; (7) \cite{Kurk2007}.}
\tablenotetext{d}{BH mass, estimated using the \mgii\ line and the TN12 prescription.}
\tablenotetext{e}{Monochromatic luminosities (\lamLlam) at observed-frame wavelengths of $\sim$3.6 and 4.5 \mic.}
\tablenotetext{f}{References for MIR photometry: (1) \Spitzer\ (\citealt{Leipski2014}; including detections by \citealt{Jiang2006_Spitzer}); (2) \WISE\ cross-match (see also \citealt{Wu2015_z6_nature} for J0100 and \citealt{Wang2015_z5_hiM} for J0306); (3) \Spitzer\ \cite[][]{Barnett2014_ULAS1120_SED}.}
\tablenotetext{g}{Obtained using \autoref{eq:mdot_ad}, \Lsptwo\ and \mbh.}
\tablenotetext{h}{Obtained using the bolometric corrections of TN12, and the \Lsptwo-based \Mdotdisk.}
\end{deluxetable*}
\capstarttrue

\end{document}